\documentclass[prl,aps,preprintnumbers,twocolumn, acknowledgement]{revtex4}

\usepackage{amsmath,amssymb,bm}
\usepackage{graphicx}
\usepackage{color}

\newcommand\q{{\mathbf{q}}}

\begin{document}
\title{Intermediate gapless phase and topological phase transition of the Kitaev model in a uniform magnetic field}
\author{Shuang Liang$^1$}
\author{Ming-Hong Jiang$^1$}
\author{Wei Chen$^{1,2}$}
\email{chenweiphy@nju.edu.cn}
\author{Jian-Xin Li$^{1,2}$}
\author{ Qiang-Hua Wang$^{1,2}$}
\affiliation{$^1$Department of Physics, Nanjing University and National Laboratory of Solid State Microstructures, Nanjing, China}
\affiliation{$^2$Collaborative Innovation Center of Advanced Microstructures, Nanjing, China 210093}

\begin{abstract}
We study the Kitaev model in a $[001]$ magnetic field employing the mean field theory in the Majorana fermion representation. The mean field Hamiltonian of the system has the Bogoliubov de-Gennes (BdG) form of a 2D superconductor. 
We discover a robust gapless regime in intermediate magnetic field for both gapless and gapped anti-ferromagnetic  Kitaev model  with $J_x=J_y$ before the system is polarized in high magnetic field. 
 A topological phase transition  connecting two gapless phases with a nodal line phase takes place at a critical magnetic field $h_{c_1}$ in this regime.
 While the nodal line phase at $h_{c_1}$  disappears when the mirror symmetry $J_x= J_y$ is broken, the nodal point gapless phase  can exist at intermediate fields even without the mirror symmetry.
 We reveal that the phase evolution of the system in the magnetic field is driven by the competition between the magnetic field and the particle-hole asymmetry of the normal state of the BdG Hamiltonian, which results in the robust intermediate gapless phase for the anti-ferromagnetic case. For ferromagnetic case, there is no intermediate phase transition before polarization. 
 The above phase diagrams are confirmed by dynamical mean field theory results.  
 \end{abstract}

\maketitle  

\section{I. Introduction}
The Kitaev spin liquid (KSL)  has attracted great interest since its proposal by Kitaev over a decade ago as it provides a minimum and exactly solvable model to study quantum spin liquid~\cite{Kitaev2006}. 
 As a spin system, the behavior of the KSL in magnetic field is of fundamental interest. The early study of Kitaev reveals that the KSL exhibits intriguing behaviors even in a perturbative weak magnetic fields, e.g., a weak magnetic field in the $[111] $ direction  can open up a gap in the spectrum of a gapless KSL and result in non-Abelian anyonic excitations~\cite{Kitaev2006},
whereas a weak magnetic field in the $[001]$ direction can close the flux gap of a gapless KSL and turn the gapless KSL to a critical spin liquid~\cite{Kitaev2011}.

 However, the study of Kitaev model in the magnetic field beyond perturbation theory is very limited until recently.  
 Motivated by recent experimental search of KSL in real materials~\cite{Wen2017, Ran2017, Wang2017, Yu2018, Ponomaryov2017}, 
 which were often conducted in a uniform magnetic field 
 to suppress the magnetic order in real materials at low temperature, 
 several theoretical groups studied the Kitaev model in uniform magnetic field, $[111]$ direction or tilted,
 employing the exact diagonalization (ED) and density matrix renormalization group (DMRG) method~\cite{Zhu2017, Gohlke2018, Trebst2018}. All these works reveal intriguing behaviors of  anti-ferromagnetic (AFM) Kitaev model in a uniform magnetic field beyond perturbation regime,
 e.g., the gapless AFM Kitaev model  exhibits a phase transition from gapped phase in weak magnetic field to a mysterious 
 gapless phase in an intermediate magnetic field before the system reaches the trivial polarized state at higher magnetic field. The origin and properties of this intermediate gapless phase remain a puzzle~\cite{Zhu2017, Gohlke2018, Trebst2018}.

 In this work,  we study the phase evolution of the KSL ground state with a $[001]$ magnetic field based on the mean field theory (MF/MFT) in the Majorana representation supplemented by the  dynamical mean field theory (DMFT) studies which includes the quantum fluctuations in time. 
This approach is naturally connected to the exact solution of the Kitaev model at zero magnetic field and keeps the generic fractionalized Majorana excitations of the KSL compared to the MFT in the widely used slave fermion representation~\cite{Liu2018}.

 The mean field Hamiltonian of the KSL in the $[001]$ magnetic field has the BdG form of a 2D superconductor~\cite{Zhang2014}. We discovered a robust gapless regime in intermediate magnetic field for both gapless and gapped AFM KSL with $J_x=J_y$ before the system is polarized in high magnetic field from the mean field theory. 
A topological phase transition connecting two gapless phases with the emergence of a nodal line takes place in the intermediate field regime at a critical magnetic field $h_{c_1}$.
 While the nodal lines at the critical field is protected by the mirror symmetry with $J_x=J_y$ and disappear at $J_x\neq J_y$ with a gap opening near the critical field, the nodal  points at $E=0$ appear due to the intrinsic particle-hole (p-h) symmetry of the BdG Hamiltonian (but not guaranteed by the symmetry) and can exist in intermediate fields even when the mirror symmetry is broken.  The above phase diagrams are confirmed by DMFT results.
 We found that the driving force for the evolution of the phase in the magnetic field is the competition between the magnetic field and the p-h asymmetry of the normal state of the mean field BdG Hamiltonian, which also results in the robust intermediate gapless phase for AFM KSL before polarizaiton.

\begin{figure}[bht]
\includegraphics[width=3cm]{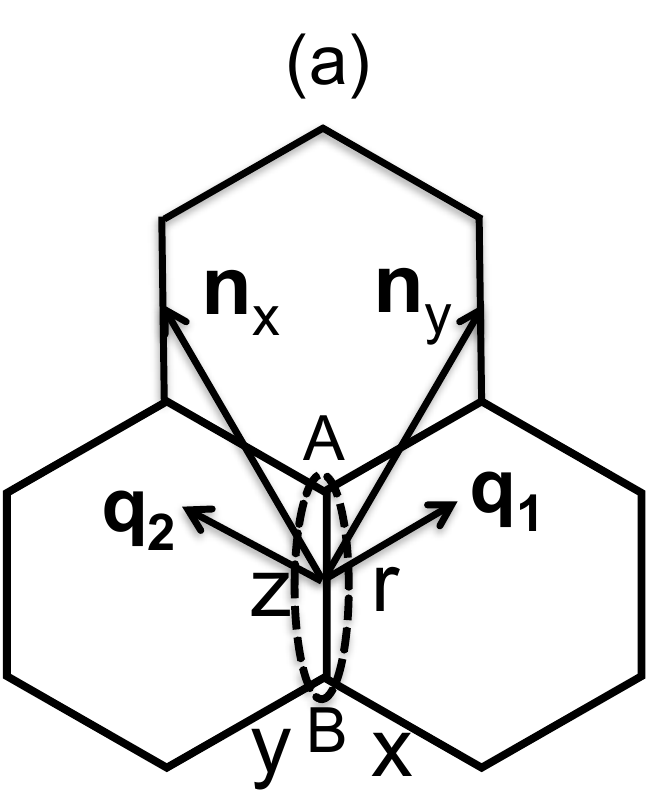}
\includegraphics[width=4.5cm]{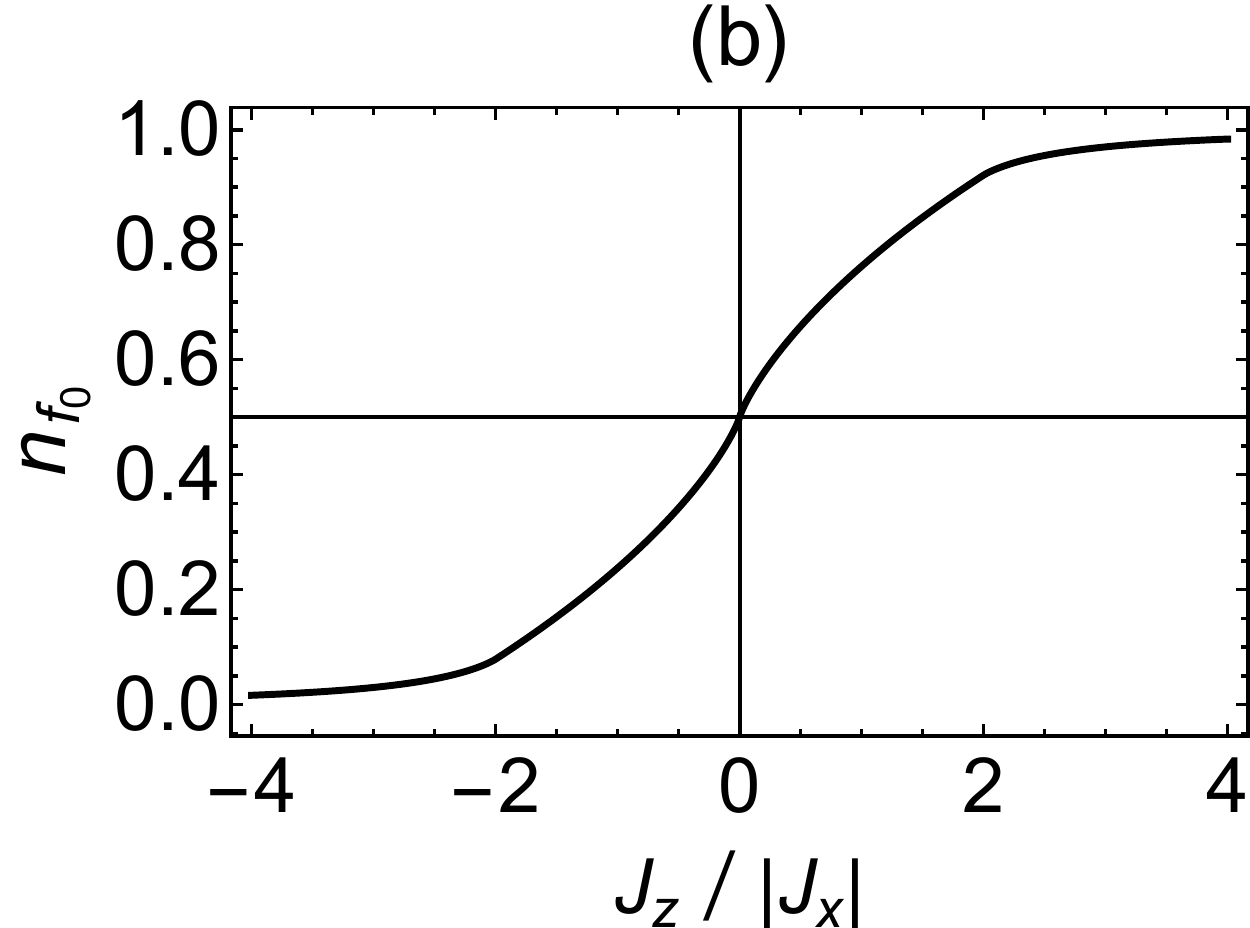}
\caption{ (a)Kitaev honeycomb lattice. The dashed oval represents a unit cell. 
$\bf{q}_1$ and $\bf{q}_2$ are unit vectors in the reciprocal lattice. (b) The occupation number of the $f$ fermion $n_{f_0}$ as a function of $J_z/|J_x|$ at the Kitaev ground state for $J_x=J_y$.
}\label{fig:lattice}
\end{figure}

\section{II. Model and Hamiltonian}
We study the KSL in a uniform magnetic field in the $[001]$ direction with the Hamiltonian 
\begin{equation}\label{eq:Hamiltonian}
H=- \sum_{\langle ij\rangle_\alpha, \alpha}J_\alpha \hat{\sigma}^\alpha_i\hat{\sigma}^\alpha_j-h\sum_i \sigma^z_i.
\end{equation}
The first term describes the pure Kitaev model on a honeycomb lattice in Fig.~\ref{fig:lattice}(a)~\cite{Kitaev2006}, 
where $\alpha=x,y,z $ and $\langle ij\rangle_\alpha$ denotes two sites sharing an $\alpha$ bond.
The second term describes the couplings of the half spin on each site with the magnetic field.

At $h=0$, the pure Kitaev model is solved by representing the half spin on each site with four Majorana fermions $\hat{c}_i, \hat{b}^x_i, \hat{b}^y_i, \hat{b}^z_i$ as $\sigma^\alpha_i=ic_i b^\alpha_i$~\cite{Kitaev2006}. This representation expands  the Hilbert space of spin to twice of the physical one  and a constraint is usually applied to narrow down the states obtained in this representation to the physical space. The pure Kitaev Hamiltonian then reads
$H_K=i\sum_{\langle ij\rangle_\alpha, \alpha} J_\alpha u_{\langle ij\rangle_\alpha} c_i c_j,$
where the bond operator $u_{\langle ij\rangle_\alpha}\equiv i b^\alpha_i b^\alpha_j$  is conserved with $u_{\langle ij\rangle_\alpha}=\pm1$. The ground state corresponds to the gauge invariant flux $W\equiv\prod_{\pi} u_{\langle ij\rangle_\alpha}$ defined on each hexagon $\pi$ to be $1$~\cite{Kitaev2006} 
and the Hamiltonian reduces to  the BdG form of a spinless p-wave superconductor when expressed in terms of complex fermions defined as $\chi_{\langle ij\rangle_\alpha}=i(b^\alpha_i+ i b^\alpha_j)/2$ and  $f_r=(c_{r,A}+ i c_{r,B})/2$~\cite{Knolle2014, Baskaran2007}, 
\begin{eqnarray}\label{eq:groundstate}
 H_0&=-&\sum_{r, \alpha=x,y} J_\alpha(f^\dag_r f_{r+n_\alpha} +f_r f_{r+n_\alpha} +h.c.)\nonumber\\
&& -\sum_r J_z(2f^\dag_r f_{r} -1)\nonumber\\
&=&\sum_{\bf q} \left[\xi^0_{\bf q} (f^\dag_{\bf q} f_{\bf q}-\frac{1}{2}) +\Delta_{\bf q} f^\dag_{\bf q} f^\dag_{-{\bf q}}+\Delta^*_{\bf q} f_{\bf q} f_{-{\bf q}}\right],\nonumber\\
 \end{eqnarray}
where $r$ represents the unit cell defined in Fig.~\ref{fig:lattice}(a) and we choose the gauge $u_{\langle ij\rangle_\alpha}=2\chi^\dag_{\langle ij\rangle_\alpha}\chi_{\langle ij\rangle_\alpha}-1=-1$, i.e, $n_{\chi_{\langle ij\rangle_\alpha}}=0$ for all the bonds in the ground state.
 $\xi^0_{\bf q}=-{\rm Re}\ S_\bold{q}$ and $\Delta_{\bf q}=i {\rm Im} \ S_\bold{q}$ is the pairing function. The Hamiltonian Eq.(\ref{eq:groundstate}) can be diagonalized by the Bogoliubov transformation $a_\bold{q} =u_\bold{q} f_\bold{q} -v_\bold{q} f_\bold{-q}^\dag$ with $u_\bold{q} =\cos \theta_\bold{q},\ v_\bold{q} =i \sin \theta_\bold{q},$ and $\tan 2\theta_\bold{q} =\frac{ \rm Im \ \it S_\bold{q}}{ \rm Re \ \it S_\bold{q}}$ ~\cite{Knolle2014}.
The spectrum is $\epsilon_0({\bf q})=\pm |S_\bold{q}|$ with $S_\bold{q}= J_z+J_x e^{i q_1}+J_y e^{i q_2}$ and does not depend on the sign of $J_\alpha$,
where  $q_1, q_2$ are the components of ${\bf q}$ in ${\bf q}_1$ and ${\bf q}_2$ directions in Fig.~\ref{fig:lattice}(a).

 The Hamiltonian $H_0$ breaks the p-h symmetry of the Kitaev model
   spontaneously as manifested by the local potential scattering of $f_r$ fermions on the $z$ bonds. The normal state of the BdG Hamiltonian  $H_0$ is then p-h asymmetric and the occupation $n_{f_0}\equiv \langle f^\dag_r f_r\rangle_0=\frac{1}{N}\sum_q |u_q|^2=\frac{1}{2}+ \frac{1}{N}\sum_q \frac{\rm Re \it S_q}{2|S_q|}$ deviates from half filling in general as shown in Fig.~\ref{fig:lattice}(b)~\cite{Chen2018}.

When the magnetic field is applied, the gauge equivalence of the ground state between the FM coupling $J_z>0$ and AFM coupling $J_z<0$ is broken, as well as the time reversal symmetry (TRS). 
The AFM KSL in a uniform magnetic field is then equivalent to a FM KSL in a staggered magnetic field on the two sublattices by a gauge transformation $J_z \to -J_z, b^z_{r,B}\to -b^z_{r,B}$ on all the $z$ bonds. However, the cases with $J_\alpha>0$ and $J_\alpha<0$ for $\alpha=x,y$ are still gauge equivalent. For the reason, we only need to consider one of the cases. We assume all the $J_\alpha< 0$ for the AFM case and all the  $J_\alpha> 0$ for the FM case in the following.

The $[001]$ magnetic field  breaks the integrability of the Hamiltonian since $u_{\langle ij\rangle_z}$ is no longer conserved,
 though, for the $x$ and $y$ bonds, $u_{\langle ij\rangle_\alpha}$ are still conserved and we set it to $-1$. Under this convention, the dimension of the remaining Hilbert space of states is the same as the physical space and there is no extra gauge constraint needed. 
  We then separate the interaction on the $z$ bonds in the Kitaev model $H_K$ 
and write the full Hamiltonian including the magnetic field as~\cite{Footnote1}
 \begin{eqnarray}\label{eq:full_hamiltonian}
       H&=&H_0 +2J_z \sum_{\bold{r}} \chi_{\bold{r}}^\dag \chi_{\bold{r}} (2 f_{\bold{r}}^\dag f_{\bold{r}} -1)\nonumber\\
       &&-2h \sum_{\bf{r}} (f_{\bf{r}}^\dag \chi_{\bf{r}} +\chi_{\bf{r}}^\dag f_{\bf{r}}),
  \end{eqnarray}
where  the sum is over all the $z$ bonds. 
The hybridization of $\chi_{\bf{r}}$ with $f_{\bf{r}}$ induced by the magnetic field flips  the sign of $u_{\bf{r}}$ and excites fluxes in the system which become mobile in the lattice. It was shown by perturbation theory that the dynamics of the flux in the above system effectively closes the flux gap and results in a power law spin correlation with distance  for gapless KSL ~\cite{Kitaev2011} instead of short-range spin correlation in pure KSL~\cite{Baskaran2007, Knolle2014}.

However, the understanding of the KSL in the magnetic field beyond perturbation theory is so far very limited.
In the following, we present an investigation of the KSL in $[001]$ magnetic field beyond perturbation based on the self-consistent MFT. 
The MFT is justified partly due to the p-h symmetry breaking of the normal state of $H_0$ so that $n_{f_0}$ deviates from half filling which suppresses the fluctuations of $\chi$ fermions in the interaction.

\begin{figure*}[tpb]
\includegraphics[width=14cm]{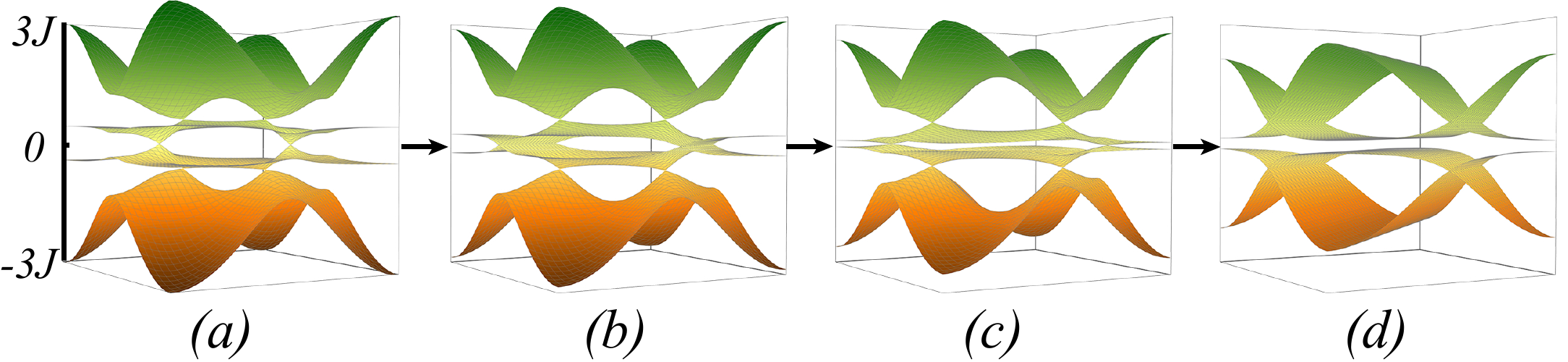}
\includegraphics[width=3.5cm]{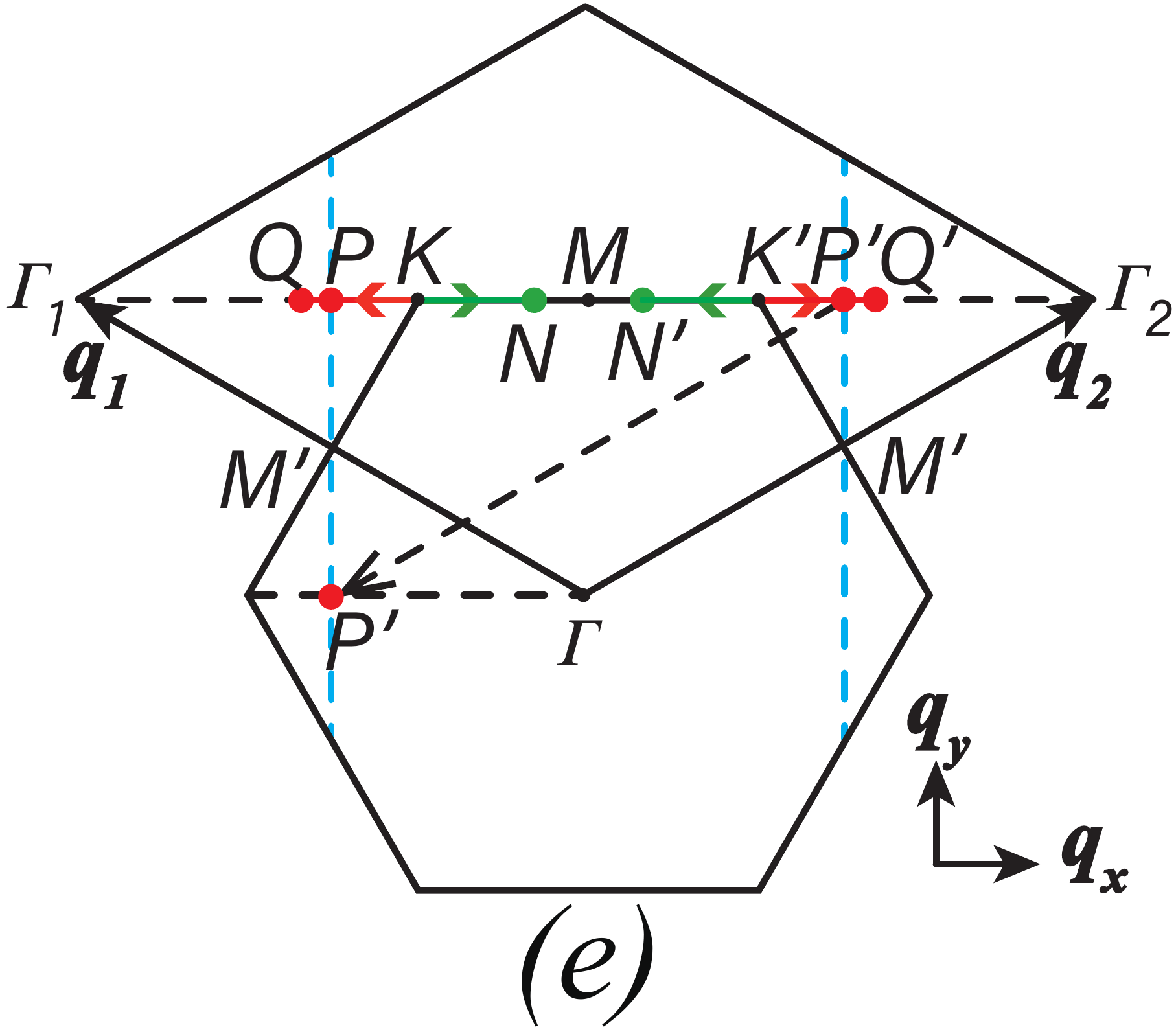}
\caption{ (a)-(d): Mean field band structure for isotropic AFM KSL with (a) $0<h<h_{c_1}$, (b)$h=h_{c_1}$, (c)$h_{c_1}<h<h_{c_2}$ and (d)$h> h_{c_2}$. (e)Evolution of Weyl points and nodal lines with the increase of magnetic field $h$ in the BZ for isotropic AFM  (red and blue lines) and FM case (green lines).}
\label{fig:AFM_band}
\end{figure*}

We decompose the interaction $H_V\equiv 2J_z \sum_{\bold{r}} \chi_{\bold{r}}^\dag \chi_{\bold{r}} (2 f_{\bold{r}}^\dag f_{\bold{r}} -1)$ in Eq.(\ref{eq:full_hamiltonian}) to the Hartree and Fock term by the standard MFT procedure
\begin{eqnarray}
      H_V^{\rm Hart}&=&4J_z \sum_{\bold{r}} n_{\chi} f_{\bold{r}}^\dag f_{\bold{r}} +4J_z \sum_{\bold{r}} (n_{f} -\frac{1}{2}) \chi_{\bold{r}}^\dag \chi_{\bold{r}},\\
       H_V^{\rm Fock} &=& -J_z \sum_{\bold{r}} m f_{\bold{r}}^\dag \chi_{\bold{r}} - J_z \sum_{\bold{r}} m^* \chi_{\bold{r}}^\dag f_{\bold{r}},
    \end{eqnarray}
 where $n_{\chi} \equiv \langle\chi_{\bold{r}}^\dag \chi_{\bold{r}}\rangle$ and $n_{f} \equiv \langle f_{\bold{r}}^\dag f_{\bold{r}}\rangle$  are the mean fields representing the average number of the $\chi_r$ and $f_r$ fermion,  and $m \equiv 4\langle\chi_{\bold{r}}^\dag f_{\bold{r}}\rangle$ is the average magnetization on the $r$-th $z$ bond. 
The $f_r$ and $\chi_r$ fermions then gain a chemical potential from the Hartree term. The Fock term results in an effective magnetic field $h_{\rm eff}=h+\frac{1}{2}J_z m$, which  is enhanced by the FM couplings and decreased by the AFM couplings.

We assume the mean fields are uniform in space~\cite{Footnote0}. The full MF Hamiltonian in the momentum space can then be written as the following $4\times 4$ matrix in the basis $\psi_{\bf{q}}=(f_{\bf{q}}, \chi_{\bf{q}}, f^\dag_{\bf{-q}}, \chi^\dag_{\bf{-q}})^T$,
\begin{equation}\label{eq:MF_hamiltonian}
      H^{\rm MF}_{\bf q} = \begin{bmatrix}
             \xi_{\bf q}  & -h_{\rm eff}    &\Delta_{\bf q}   & 0 \\
            -h_{\rm eff}^*    & \xi_\chi    &0    & 0 \\
           \Delta^*_{\bf q}     & 0    &  - \xi_{\bf q}    &h_{\rm eff}^*\\
            0    &0    & h_{\rm eff}    & -\xi_\chi
            \end{bmatrix},
    \end{equation}
where $\xi_{\bf q}=-{\rm Re}\ S_\bold{q}+2 J_z n_{\chi} $ and $\xi_\chi= 2J_z(n_f-1/2)$ are the normal state spectrum of $f$ and $\chi$ fermions respectively. We note that $\xi_\chi>0$ for both AFM and FM KSL, which results in a finite energy cost for flux excitation and suppresses the fluctuations of $\chi$ fermions.

The above Hamiltonian can be written as $H^{\rm MF}_{\bf q}=\tilde{H}_N \tau_z+ \tilde{\Delta}\tau_y$, where $\tilde{H}_N$ is the upper diagonal $2\times 2$ block describing the normal state, $\tilde{\Delta}$ is the upper off-diagonal $2\times 2$ matrix describing the pairing and $\tau_z$ and $\tau_y$ act on the Nambu space. Though the p-h symmetry of the normal state $\tilde{H}_N$ is broken, the BdG Hamiltonian has an intrinsic p-h symmetry ${\cal C}H^{\rm MF}_{\bf q}{\cal C}^{-1}=-H^{\rm MF}_{\bf q}$  under ${\cal C}=\tau_y K$ (K is the complex conjugate operator) due to the p-h redundancy of the Nambu space which results in a p-h symmetric spectrum~\cite{Zhang2014}.

\section{III. Results from the mean field theory}

\subsection{III.1. Mirror symmetric case $J_x=J_y$}
The MF Hamiltonian Eq.(\ref{eq:MF_hamiltonian}) is solved self-consistently. The $[001]$ magnetic field breaks the ${C}_3$ rotational symmetry  of the  lattice, however, the mirror symmetry with respect to the ${\bf q}_x$ and ${\bf q}_y$ axis crossing the $\Gamma$ point in Brilliouin zone (BZ) is preserved  for $J_x=J_y$. We study this mirror symmetric case at first.

\subsubsection{1. AFM case}

{\it (a)Gapless AFM case.} We first study the gapless AFM case with mirror symmetry.  For the typical isotropic case $J_x=J_y=J_z<0$, the evolution of the four bands  from the MF Hamiltonian with magnetic field is shown in Fig.~\ref{fig:AFM_band}(a)-(d).
It  reveals that the two Weyl points (in some references called Dirac points) $K$ and $K'$ at $E=0$
 and $h=0$ persist and move along the $K\Gamma_1$ and $K'\Gamma_2$ direction all the way to $Q$ and $Q'$ indicated by the red lines in Fig.~\ref{fig:AFM_band}(e) as $h$ increases from $0$ to $h_{c_2}$, except that at some critical magnetic field $h=h_{c_1}<h_{c_2}$, the Weyl points moving to $P=(\pi/2, 3\pi/2)$ and $P'=(3\pi/2, \pi/2)$ in the ${\bf q}_x$ direction turn to two nodal lines crossing $PP'$ in the ${\bf q}_y$ direction (blue dashed lines  in Fig.~\ref{fig:AFM_band}(e)) as shown in Fig.~\ref{fig:AFM_band}(b). At $Q$ and $Q'$, the two Weyl points disappear simutaneously and a gap opens at  $\Gamma$ point as shown in the Fig.~\ref{fig:AFM_band}(d). The band structure then indicates two phase transitions at $h_{c_1}$ and $h_{c_2}$ respectively.

\begin{figure}[tpb]
\includegraphics[width=8cm]{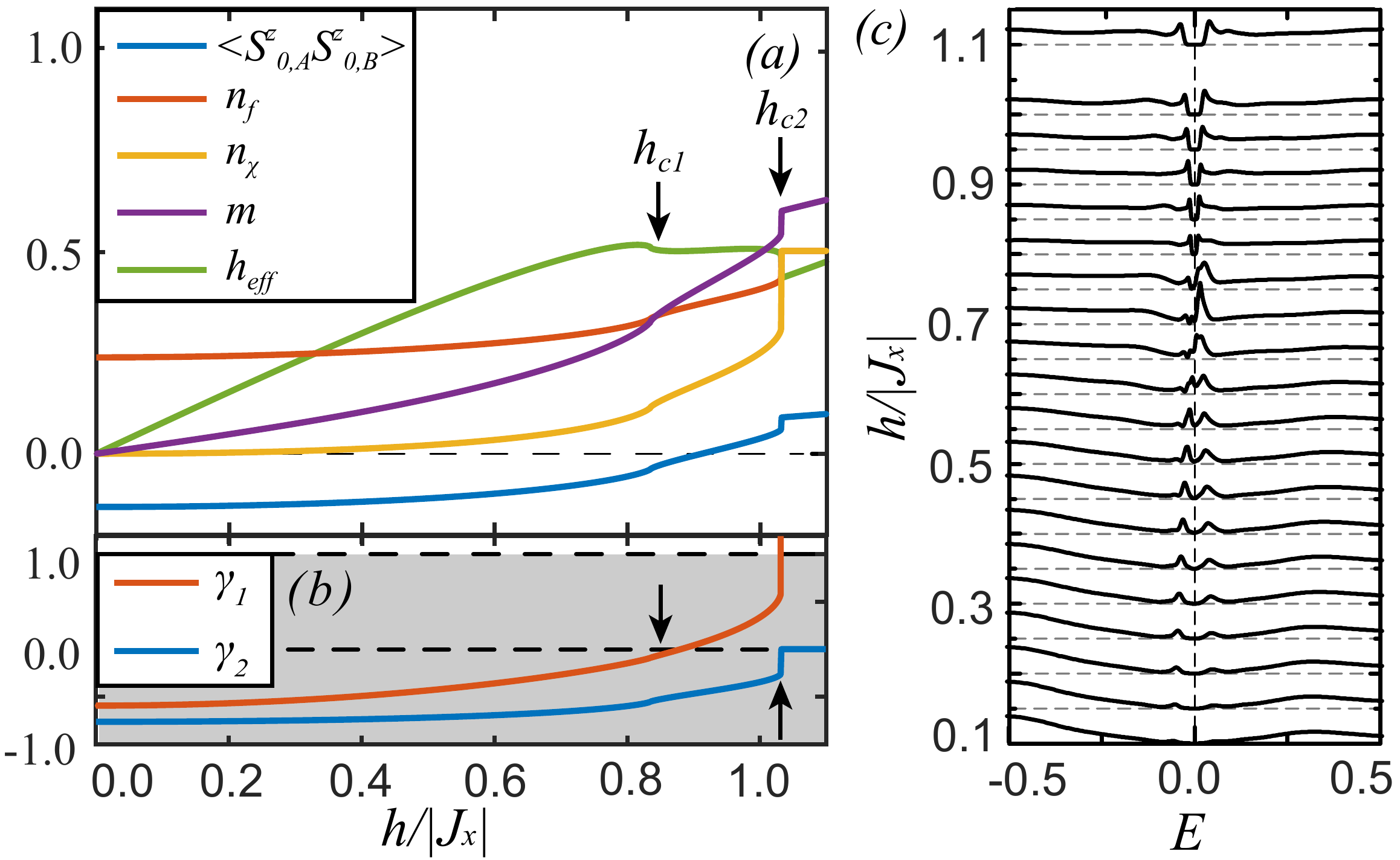};                
\caption{ (a) and (b): The mean field parameters, spin-correlation function on the $z$ bonds and the parameters $\gamma_1$ and $\gamma_2$ as a function of $h$ obtained from the MFT for isotropic AFM KSL. (c)The evolution of the low energy density of state of the matter fermion quasiparticles with the magnetic field obtained from the DMFT. The horizontal dashed lines locate the zero point of the density of states for each curve. The vertical axis labels the magnetic field $h/|J_x|$ for the corresponding curve. The unit of energy $E$ is $|J_x|$.}\label{fig:AFM_results}
\end{figure}

The plots of the mean field parameters as a function of $h$ in Fig.~\ref{fig:AFM_results}(a) also reveal phase transitions at $h_{c_1}$ and $h_{c_2}$. The magnetization jumps discontinuously at $h_{c_2}$ which indicates a first order phase transition to partial-polarized state. The occupation of both $f$ and $\chi$ fermions increases with $h$ and jumps to $1/2$ at the transition point $h_{c_2}$. The high magnetic field then tends to restore the p-h symmetry of the normal state $\tilde{H}_N$ because the maximum magnetization requires half filling of both $n_f$ and $n_\chi$. We will show below that this tendency drives the evolution of the phase and phase transition with the increase of the magnetic field.

The mean field parameters at $h_{c_1}$ only show small kinks (resulting in a peak in magnetic susceptibility) as shown in Fig.~\ref{fig:AFM_results}(a), indicating a continuous phase transition at $h_{c_1}$. We will focus on this phase transition in the following.

From the mean field Hamiltonian Eq.(\ref{eq:MF_hamiltonian}), the zero energy solution appears when both the pairing function $\Delta_{\bf q}$ and the determinant of $\tilde{H}_N$ is zero, i.e.,
$\tilde{H}_N$ has a zero eigenenergy,
\begin{equation}\label{eq:Weyl_location}
         \left\{
             \begin{array}{ll}
              \Delta_{\bf q}\propto{\rm Im} \ S_\bold{q}  =0,\\
             \xi_{\bf{q}} \xi_{\chi} - |h_{\rm eff}|^2 =0.
             \end{array}
          \right.
       \end{equation}
 The intrinsic p-h symmetry of the BdG Hamiltonian guarantees that any zero energy point is at least two fold degenerate. 
 
 In the case $J_x=J_y$, Eq.(\ref{eq:Weyl_location}) has the following solutions
\begin{equation}\label{eq:solution_1}
      \left\{
             \begin{array}{ll}
              q_1 = -q_2 \\
              \cos q_1 = \cos q_2 = \gamma_1
            \end{array}
           \right. , \rm and \quad
      \left\{
             \begin{array}{ll}
              q_1 = q_2 \pm \pi\\
              \cos q_1 = \cos q_2 =\gamma_1            \end{array}
           \right. ,
      \end{equation}
      where $\gamma_1=(n_\chi-\frac{1}{2})J_z/J_x+\frac{|h_{\rm eff}|^2}{4J_x J_z(\frac{1}{2}-n_f)}$. The first  solution requires $|\gamma_1| \leq1$ and gives the locations of two Weyl points at $q^*_1=-q^*_2, \ q^*_1=\pm \arccos\gamma_1$. We can see that the Weyl points move along the ${\bf q}_x$ direction in the BZ as shown in Fig.\ref{fig:AFM_band}(e).
   Whereas the second solution exists if and only if $\gamma_1 =0$ and $J_x=J_y$. In this case, the solution becomes two nodal lines at $q_1=q_2\pm \pi$. This takes place at the critical magnetic field $h=h_{c_1}$.

For AFM KSL, both $n_f$ and $n_\chi$ increases  monotonically with $h$ from less than $1/2$ at $h=0$ until at $h=h_{c_2}$, both jumps to $1/2$.
 The effective magnetic field $h_{\rm eff}$ also increases with $h$ for $h<h_{c_2}$ as shown in Fig.~\ref{fig:AFM_results}(a). For the reason, $\gamma_1$ increases monotonically with $h$ for gapless KSL from $-1\leq \gamma_1=-J_z/2J_x<0$ at $h=0$ until at $h=h_{c_2}$, $\gamma_1$ jumps from less than $1$ ($\gamma_1=1$ corresponds to the $\Gamma_1$ and $\Gamma_2$ point)  to $-\frac{|h_{\rm eff}|^2}{4J_zJ_x(n_f-1/2)}\to +\infty$. The Weyl point solution at $E=0$ for gapless AFM KSL with $J_x=J_y$  are then expected to exist  all the way from $h=0$ to $h=h_{c_2}$ except that at $h=h_{c_1}$, $\gamma_1=0$, and the Weyl points turn to two nodal lines.

The variation of $\gamma_1$ with $h$ for $J_x=J_y=J_z<0$ in Fig.~\ref{fig:AFM_results}(b) and the corresponding band structure in Fig.~\ref{fig:AFM_band}(a)-(d) confirmed the above analysis. We can see that the driving force for the phase evolution of the AFM KSL with the magnetic field is the competition between the magnetic field and the p-h asymmetry of the normal state so that $n_f$ and $n_\chi$ both tends to increase to half filling as $h$ increases.

To have a better understanding of the topological properties of the Weyl points at $E=0$ and the phase transition at $h=h_{c_1}$, we obtain an effective $2\times 2$ Hamiltonian near the Weyl points for the two bands touching at the Weyl points by projecting the MF Hamiltonian Eq.(\ref{eq:MF_hamiltonian}) with the projecting operator containing only the two degenerate states at each Weyl point, i.e., $H_{2\times2}=P^\dag H_{\bf q}^{\rm MF} P$ with $P=|\Psi_{01}\rangle\langle \Psi_{01}|+|\Psi_{02}\rangle\langle \Psi_{02}|$, where $|\psi_{01}\rangle$ and $|\psi_{02}\rangle$ are the two degenerate  eigenstates at each Weyl point. This is justified because near the Weyl points at $E=0$, the mixing from the highest and lowest band to the eigenstates is small and does not affect the topological properties of the Weyl points.
As shown in the appendix, the effective $2\times 2$ Hamiltonian for the $J_x=J_y$ case after expanding to linear order of $\delta {\bf q}\equiv {\bf q-q}^*$ near the Weyl points ${\bf q}^*$ at $E=0$  is
\begin{equation}
H_{2\times2}=-\frac{2J_{x} |h_{\rm eff}|^2}{|h_{\rm eff}|^2+\xi^2_{\bf{q^*}}}
             \begin{bmatrix}
             0 & \tilde{v}_x \delta q_x -i \tilde{v}_y \delta q_y \\
               \tilde{v}_x \delta q_x+i \tilde{v}_y \delta q_y  & 0
            \end{bmatrix},        
    \end{equation}
where $q_x=(q_2-q_1)/2,  q_y=(q_2+q_1)/2$, $\tilde{v}_x=\sin q^*_1=\pm\sqrt{1-\gamma^2_1}$, $\tilde{v}_y =\cos q^*_1=\gamma_1$. The basis composed of the two degenerate states for $H_{2\times2}$ are properly chosen as shown in the Appendix. 
The dispersion near the Weyl point is then $E^\pm_{\bf{q}}=\pm\frac{2J |h_{\rm eff}|^2}{|h_{\rm eff}|^2+\xi^2_{\bf{q^*}}}\sqrt{\tilde{v}^2_x \delta q^2_x+\tilde{v}^2_y \delta q^2_y}$. The upper and lower band has chirality $\pm1$ respectively.

The eigenvectors of the effective Hamiltonian near each Weyl point are two pseudospin vectors $|\tilde{\psi}\rangle_\pm=\frac{1}{\sqrt{2}}(1, \pm e^{i\rm \tilde{\theta}_{\bf q}} )^T$, where $\tilde{\theta}_{\bf q}$ is the angle between ${\bf \tilde{q}}=(\tilde{v}_x \delta q_x, \tilde{v}_y \delta q_y)$ and the ${\bf q}_x$ axis,
and rotates a full circle when $\delta {\bf q}$ rotates a circle around the Weyl point, corresponding to a winding number $+1$ or $-1$ of the eigenvectors~\cite{Park2011}. 
For a given Weyl point,  $\tilde{v}_y = \gamma_1$ changes sign at $h=h_{c_1}$ as shown in Fig.~\ref{fig:AFM_results}(b) whereas $\tilde{v}_x$ does not, so the winding number also changes sign resulting in a  topological phase transition at $h_{c_1}$. Since $\tilde{v}_x$  has opposite signs at the two Weyl points whereas $\tilde{v}_y$ is the same at the two points, the two Weyl points have opposite winding number.

There are two other pairs of Weyl points at finite energy $\epsilon_\bold{q^*} \neq 0$ at $h<h_{c_2}$ as shown in Fig.~\ref{fig:AFM_band}(a)-(d). 
 The Weyl points at $\epsilon_\bold{q^*} \neq 0$ satisfy:
    \begin{equation}\label{eq:condition_2}
        \left\{
             \begin{array}{ll}
              {\rm  Im} \ S_\bold{q} =0,\\
              \xi_\bold{q} + \xi_\chi =0.
             \end{array}
         \right.
    \end{equation}
    These Weyl points require the pairing function $\Delta_{\bf q}=0$ and at the same time, the two eigenenergies of  $\tilde{H}_N$ are opposite to each other.
    Due to the p-h symmetry of the bands, these solutions form Weyl points at the corresponding $\pm E_N$.  
    
  In the case $J_x=J_y$, the solution of Eq.(\ref{eq:condition_2}) has the same form of Eq. (\ref{eq:solution_1}) except that $\gamma_1$ is replace by $\gamma_2=(n_{\chi} + n_{f} -1)J_z/J_x$.   Similar analysis to the above applies to the Weyl points at $\pm E_N$ only with $\gamma_1$ replaced by $\gamma_2=(n_{\chi} + n_{f} -1)J_z/J_x$. Since $\gamma_2 =(n_{\chi} + n_{f} -1)J_z/J_x\leq0$ at $h\leq h_{c_2}$, i.e., it does not change sign at $h<h_{c_2}$, the topological characters of these Weyl points do not change at $h< h_{c_2}$.

{\it (b) Gapped AFM case.} For the gapped AFM  case with $J_x=J_y$, $\gamma_1=-J_z/2J_x<-1$ at $h=0$ so the spectrum is gapped. As $h$ increases, $\gamma_1$ increases from the above analysis until at the trivial polarized phase $\gamma_1\to +\infty$. 
There then must exist an intermediate range of $h$ where $-1\leq\gamma_1\leq 1$ and the gapped KSL becomes gapless from the mean field Hamiltonian. This is confirmed by the band structure of the case $J_z=2.5J_x=2.5J_y<0$ shown in the appendix. There also exists an intermediate magnetic field $h_{c_1}$ where $\gamma_1=0$ and a topological phase transition takes place between two gapless phases. 

From the above analysis, we see that for the AFM KSL with $J_x=J_y$, either gapless or gapped, since $\gamma_1<0$ at $h=0$ and $\gamma_1\to +\infty$ at $h\to h_{c_2}$, there must exist a gapless regime at intermediate magnetic field and a topological phase transition corresponding to $\gamma_1=0$ before the system reaches the polarized state at high magnetic field.

\subsubsection{2. FM case}

\begin{figure}[tpb]
\includegraphics[width=7.5cm]{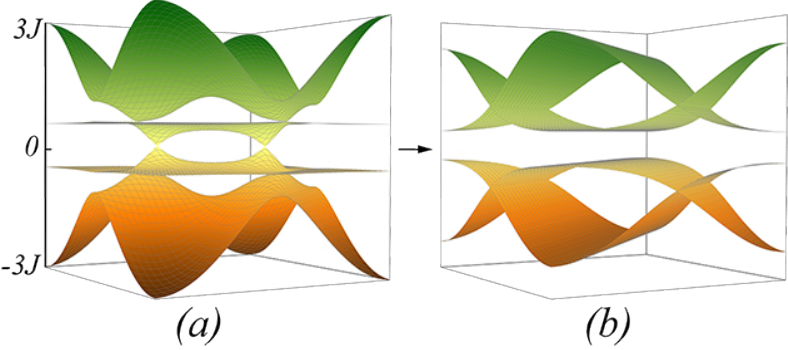}               
\caption{Band structure for the isotropic FM case from the MFT. (a) $0<h<h_c^{\rm FM}$. (b)$h> h_c^{\rm FM}$.}\label{fig:FM_band}
\end{figure}

\begin{figure}[tpb]
\includegraphics[width=8cm]{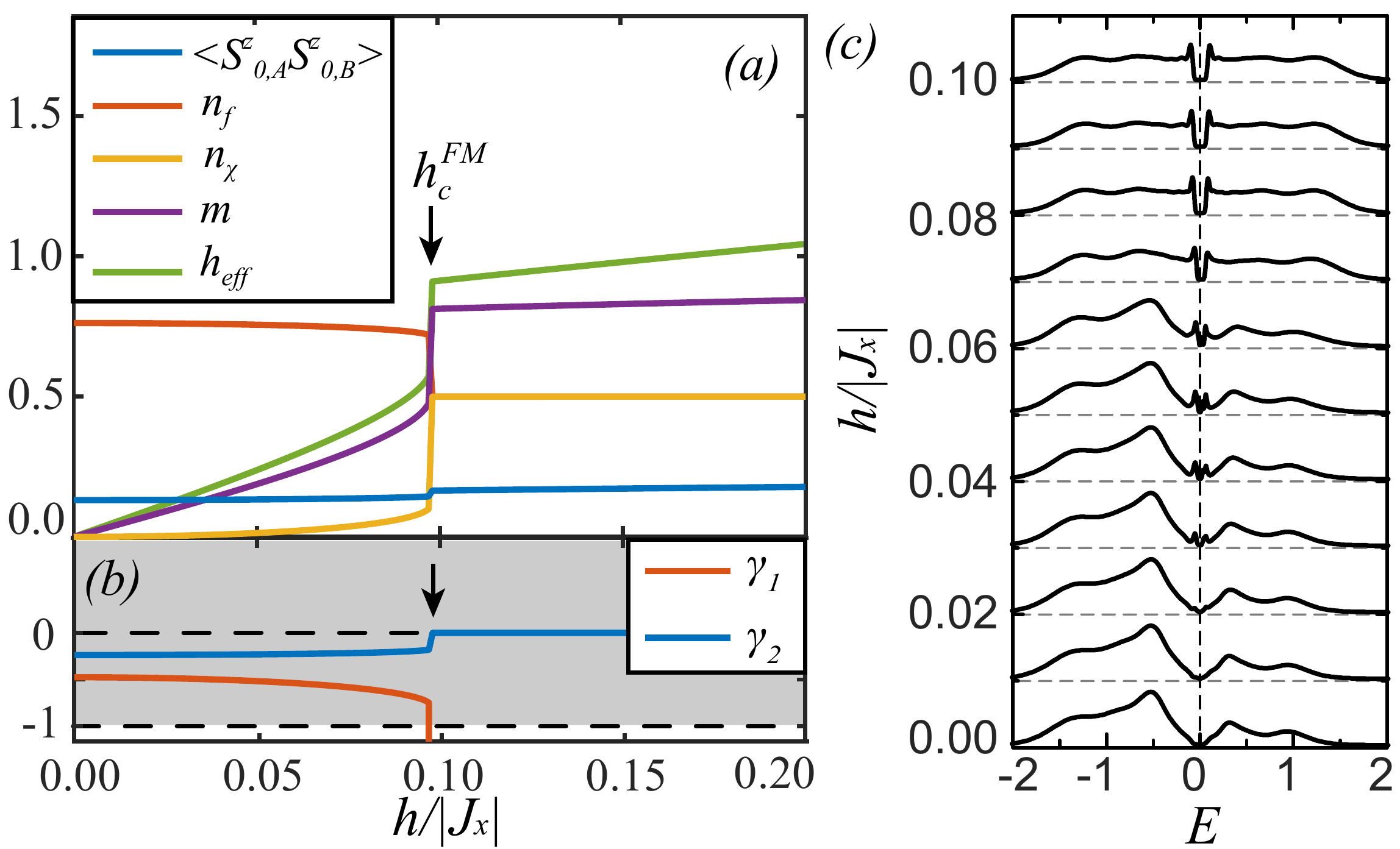}               
\caption{(a) and (b): The mean field parameters, spin-correlation function on the $z$ bonds and the parameters $\gamma_1$ and $\gamma_2$ as a function of $h$ obtained from the MFT for isotropic FM case. (c)The evolution of the low energy density of state of the matter fermion quasiparticles with the magnetic field obtained from the DMFT for the isotropic FM case. The labels of the curves are the same as for the AFM case in Fig.\ref{fig:AFM_results}(c).}\label{fig:FM_results}
\end{figure}

 For the FM case with $J_x=J_y$, $n_f>1/2, n_{\chi}<1/2$  and $\gamma_1=-J_z/2J_x<0$ at $h=0$. As $h$ increases from zero, $\gamma_1$ decreases for both gapless and gapped KSL until at some critical magnetic field $h=h_c^{\rm FM}$, $\gamma_1\to -\frac{|h_{\rm eff}|^2}{4J_zJ_x(n_f-1/2)} \to -\infty$.  
 For the reason, there is no intermediate phase transition from the MFT before reaching the trivial polarized phase at $h=h_c^{\rm FM}$ for FM KSL. 
This is confirmed by the band structure of the typical FM case $J_x=J_y=J_z>0$ shown in  Fig.\ref{fig:FM_band}. The corresponding  
  mean field parameters and $\gamma_1$ are shown in Fig.\ref{fig:FM_results}(a) and (b).   At $h=h_c^{\rm FM}$, the system experience a first order phase transition to the gapped polarized state at $h=h_c^{\rm FM}$ for the gapless KSL. Whereas for gapped FM KSL, the transition to the gapped polarized phase is continuous.
  
  The critical magnetic field $h^{\rm FM}_{c}$ to reach polarized phase for the FM KSL is much lower than its counterpart $h_{c_2}$ for the AFM KSL. This is because the FM Kitaev coupling enhances the effective magnetic field whereas the AFM coupling decreases $h_{\rm eff}$.

\begin{figure*}[tpb]
\includegraphics[width=14cm]{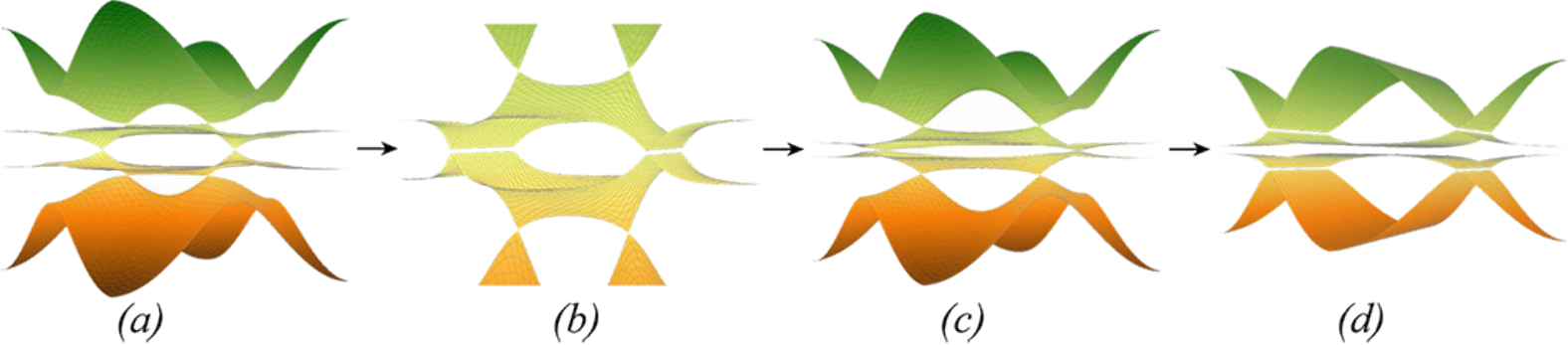}               
\caption{(a)-(d): Mean field band structure for the mirror asymmetric AFM case $J_x=J_z, J_y=0.95J_x$. (a) $0<h<h'_{c_1}$, (b)$h'_{c_1}<h<h''_{c_1}$, (c)$h''_{c_1}<h<h_{c_2}$ and (d)$h>h_{c_2}.$}\label{fig:ASM_band}
\end{figure*}

\begin{figure}[tpb]
\includegraphics[width=8.5cm]{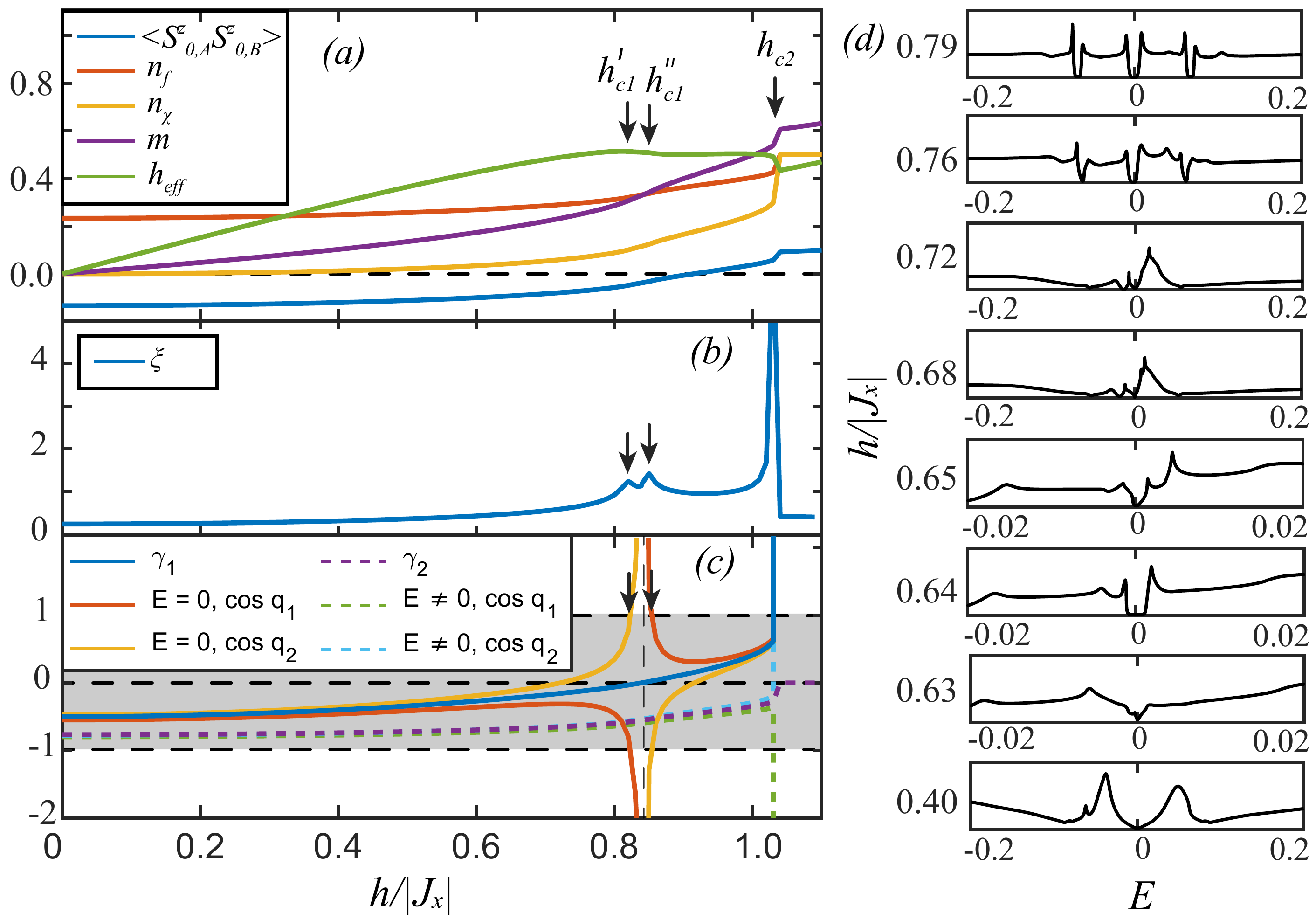}               
\caption{(a): The mean field parameters and the spin-correlation function on the $z$ bonds obtained from the MFT for the mirror asymmetric AFM case $J_x=J_z, J_y=0.95J_x$. (b)The magnetic susceptibility as a function of $h$ for the corresponding case. (c)The parameters $\gamma_1$ and $\gamma_2$, and  ${\rm cos q_1}$ and $\rm{ cos q_2}$ corresponding to the Weyl point solutions as a function of $h$. The solid lines are for zero energy Weyl points in Eq. (\ref{eq:ASM_solution}) and the dashed lines are for Weyl points at $E\neq 0$. (d)The evolution of the low energy density of state of the matter fermion quasiparticles with the magnetic field obtained from the DMFT for the same case. The labels of the curves are the same as for the AFM case in Fig.\ref{fig:AFM_results}(c).}\label{fig:ASM_results}
\end{figure}

\subsubsection{3. Comparison with the DMFT results}
 To check the reliability of the above results from the MFT, we carried out the DMFT, which includes the quantum fluctuations in time, for the Hamiltonian Eq.(\ref{eq:full_hamiltonian}) in different cases. 
 
 The evolution of the  low energy density of states (DOS) of the matter fermion quasiparticles as a function of the  magnetic field for the corresponding AFM case and FM case is shown in Fig.~\ref{fig:AFM_results}(c) and
 Fig.~\ref{fig:FM_results}(c) respectively. For the isotropic AFM case, the DMFT result in Fig.~\ref{fig:AFM_results}(c) reveals that the low energy DOS of the matter fermion quasiparticles changes from linearly vanishing with energy at low magnetic field to a finite value at some intermediate magnetic field and then to linearly vanishing again with the increase of $h$. Finally at high magnetic field, the low energy DOS vanishes within an energy gap  at $E=0$. This is qualitatively consistent with the phase evolution of the isotropic AFM case in the magnetic field obtained from the MFT in the previous subsection, i.e., from a gapless phase to a nodal line phase and then to a gapless phase again with the increase of the magnetic field and finally a gap opens at $E=0$ at high magnetic field. The main difference  between the MFT and dynamical MFT is that the critical magnetic field for the intermediate phase transition is lower from the dynamical MFT and the finite DOS at $E=0$ seems to exist in a range of intermediate magnetic field instead of only at the critical field $h_{c_1}$. 
 
The phase evolution obtained from the DMFT  for  the FM case  in Fig.~\ref{fig:FM_results}(c) and the gapped AFM case in Fig.~\ref{fig:gap_afm_para}(c) in the appendix also agrees qualitatively with the MFT results for the corresponding case.

\subsection{III.2. Mirror asymmetric case $J_x\neq J_y$}

 In the case $J_x\neq J_y$, the mirror symmetry in the ${\bf q}_x$ and ${\bf q}_y$ direction is broken. The zero energy nodal line solution for the AFM case at  $\gamma_{1}=0$ from the MFT  disappears. But the Weyl point solution at $E=0$ can still exist for the mean field Hamiltonian. 
  Let $\tilde{\gamma_1} \equiv 2J_x \gamma_1$,
         the zero energy solution from Eq. (\ref{eq:Weyl_location}) then becomes:
        \begin{align}
          \begin{cases}
          \cos q_1  = \frac{\tilde{\gamma}_1^2 - J_y^2 +J_x^2}{2 J_x \tilde{\gamma}_1},\ & \sin q_1  = \pm \sqrt{1-\cos^2 q_1 }, \\
          \cos q_2  = \frac{\tilde{\gamma}_1^2 + J_y^2 -J_x^2}{2 J_y \tilde{\gamma}_1},\ & \sin q_2  = \mp  \sqrt{1-\cos^2 q_2 }.
          \end{cases}\label{eq:ASM_solution}
         \end{align}
The Weyl point exists when both $|\cos q_1|\leq 1$ and $|\cos q_2|\leq 1$. At $\gamma_1\to 0$, $\tilde{\gamma_1}\to 0$ as well,  both $\cos q_1$ and $\cos q_2$  then goes to infinity. The zero energy solution at $\gamma_1=0$ for the mirror symmetric AFM case  disappear in the case $J_x\neq J_y$ and a gap opens at $E=0$ near the neighborhood of $h_{c_1}$ ($\gamma_1=0$ at $h=h_{c_1}$).  
Fig.~\ref{fig:ASM_band} shows the  band structure from the MFT for a typical AFM gapless mirror asymmetric KSL in the magnetic field with $J_x=J_z<0, J_y=0.95J_x$. 
The corresponding mean field parameters and the zero energy solutions are shown in Fig.~\ref{fig:ASM_results}(a)-(c). From both figures, we see that a gap opens at $h'_{c_1}<h< h''_{c_1}$, where $h'_{c_1}<h_{c_1}< h''_{c_1}$.
The range $h''_{c_1}-h'_{c_1}\sim |J_x-J_y|$. For not very large $|J_x-J_y|$ compared to $|J_x|$, there still exists an intermediate gapless regime at $h''_{c_1}<h<h_{c_2}$ as shown in Fig.~\ref{fig:ASM_band}(c) and Fig.~\ref{fig:ASM_results}(c). This results in two successive phase transitions at $h'_{c_1}$ and $h'_{c_2}$ as shown in Fig.~\ref{fig:ASM_results}(a) and (b). 

We also checked the DMFT results for the case $J_x=J_z<0, J_y=0.95J_x$. As shown in Fig.~\ref{fig:ASM_results}(d), the low energy DOS of the matter fermion quasiparticles near $E=0$ indicates a gapped phase at intermediate magnetic field around $h/|J_x|\sim 0.63-0.65$ and  gapless phases at around  $0<h/|J_x|\leq 0.63$ and $0.65 \leq h/|J_x|\leq 0.75$. At  $h/|J_x|>0.76$, the system reaches the gapped polarized state. This is qualitatively consistent with the phase evolution from the MFT shown in Fig.~\ref{fig:ASM_band} and Fig.~\ref{fig:ASM_results}(c).

The two gapless regime at $0<h<h'_{c_1}$ and $h''_{c_1}<h< h_{c_2}$ has opposite winding number around the Weyl points as shown in the appendix. The gapped phase at 
$h'_{c_1}<h< h''_{c_1}$ due to the mirror symmetry breaking is trivial with total Chern number zero for the filled bands $E<0$. This is in contrast to the gapped phase resulting from a small tilting of the magnetic field from the $[001]$ direction, which is topologically nontrivial with a Chern number ${\cal C}=\pm 1$~\cite{Kitaev2006} and results in gapless edge modes and quantum thermal Hall effect~\cite{Nomura2012, Fujimoto2013}.

\section{IV. Discussions and conclusions}

 The mean field state in the fractionalized majorana representation is by construction a quantum spin liquid state before reaching the trivial polarized phase at high magnetic field. Due to the pairing term of the matter fermions, the spin liquid phase of the mean field Hamiltonian in this work is  different from the $U(1)$ spin liquid phase obtained from the mean field Hamiltonian of the $K-\Gamma$ model in the magnetic field in Ref.\cite{Liu2018} where the number of spinons is conserved. The spin liquid phase  from the MFT here is more likely a $Z_2$ spin liquid due to the pairing of the matter fermions, which, however, is also different from the $Z_2$ spin liquid phase of a pure Kitaev model, where a well defined finite flux gap exists while in the $[001]$ magnetic field the flux gap is closed. The nature of the states in the magnetic field in this work deserves further study in future  work.

 The gapless phase with Dirac points also persists in the $K-\Gamma$ model with the magnetic field in selected directions as studied in Ref.\cite{Liu2018}. There the Dirac cones are protected by the $C_{3v}$ lattice symmetry or the emergent sublattice-spin rotation symmetry due to the spin-orbit coupling, while in our case, the Weyl points appear due to the  intrinsic p-h symmetry 
of the  BdG Hamiltonian which guarantees that any zero energy point is two fold degenerate. 
The appearance of the zero energy solution is however accidental.

 We note that 
 the phase transition at $h_{c_1}$ for AFM KSL with $J_x=J_y$ is not simply equivalent to a sign change of the effective Kitaev coupling $J_z$ in pure Kitaev model as suggested in \cite{Nasu2018}. In fact, the spin correlation function on the $z$ bond changes sign for the AFM KSL at some $h$ between $h_{c_1}$ and $h_{c_2}$ instead of $h_{c_1}$ as shown in Fig.\ref{fig:AFM_results}(a). 
 Besides,  the normal state of the pure Kitaev BdG Hamiltonian at $J_z=0$ is p-h symmetric with $n_f=n_\chi=1/2$, 
whereas  this p-h symmetry of the normal state with finite $J_z$ is restored at $h=h_{c_2}$ instead of $h_{c_1}$.

The gapless phase was also discovered  for the AFM Kitaev model in an intermediate magnetic field in the $[111]$ or tilted directions in recent numerical works applying the DMRG or exact diagonalization method~\cite{Zhu2017, Gohlke2018, Trebst2018}.  For the  Kitaev model in an arbitrarily directed magnetic field, the bond operator $u_{\langle ij\rangle_\alpha}$ gains a dynamics on all the bonds and the gauge convention $u_{\langle ij\rangle_{x,y}}=-1$ used in this work does not apply. A widely used constraint in this case is the local constraint $D_j\equiv b_j^x b_j^y b_j^z c=1$  introduced in Ref~\cite{Kitaev2006}. The comparison of the mean field results in the $[111]$ and tilted magnetic field with previous numerical works will be left for a future study.

In conclusion, we studied the phase evolution of the Kitaev model in $[001]$ magnetic field based on MFT and DMFT. In the mirror symmetric case $J_x=J_y$, we discovered a robust gapless phase at intermediate magnetic field and a topological phase transition between two gapless phases connected by a nodal line phase for both gapless and gapped AFM KSL before reaching the trivial polarized state at high magnetic field. Whereas for FM KSL, there is no intermediate phase transition before polarization. While the nodal line phase is protected by mirror symmetry and disappears at $J_x\neq J_y$, the intermediate gapless phase can still exist even without mirror symmetry. We discovered that the phase evolution is driven by the competition between the magnetic field and the p-h asymmetry of the normal state of the mean field BdG Hamiltonian, which also results in the robust intermediate gapless phase. The phase evolution of the system from the MFT is qualitatively consistent with the results from the DMFT.

 \begin{acknowledgements}
{\it Acknowledgement.} We thank Xian-Gang Wan, Hai-Jun Zhang, Zheng-Xin Liu, Itamar Kimchi, Zhao-Yang Dong, Zhao-Long Gu and Kai Li for helpful discussion. This work is supported by the NSF of China under Grant No.11504195(WC), No.11774152(JXL), No.11574134(QHW), the National Key Projects for Research and Development of China under Grant No. 2016YFA0300401(JXL and QHW) and Jiangsu Province Educational department(WC).
\end{acknowledgements}

 {\it Note Added:} During the writing of this paper, we noticed a recent preprint on arXiv.\cite{Nasu2018} with partial overlapping results of ours and the work in Ref.~\cite{Trebst2018} on a similar problem studied by exact diagonalization.

\section{Appendix}

\section{I. Effective  $2 \times 2$ Hamiltonian near the Weyl points}

      We obtain an effective $2 \times 2$ Hamiltonian near the Weyl points for the two bands touching at the Weyl points by projecting the mean field Hamiltonian with the projecting operator containing only the two degenerate states at the Weyl points, i.e., $H_{2\times2}=P^\dag H_{\bf q}^{\rm MF} P$ with $P=|\Psi_{01}\rangle\langle \Psi_{01}|+|\Psi_{02}\rangle\langle \Psi_{02}|$, where $|\psi_{01}\rangle$ and $|\psi_{02}\rangle$ are the two degenerate  eigenstates at each Weyl point.

      At Weyl points ${\bf q}^*$, the MF Hamiltonian has a block diagonal form since $\Delta_{\bf q^*}=0$:
      \begin{equation}
      H_{\q^*} = \begin{bmatrix}
            \xi_{\q^*}    & -h_{\rm eff}    &0              & 0 \\
            -h_{\rm eff}^*     & \xi_\chi    &0              & 0 \\
            0              & 0           & -\xi_{\q^*}    &h_{\rm eff}^*\\
            0              & 0           & h_{\rm eff}       & -\xi_\chi
            \end{bmatrix}.
      \end{equation}

     The  two eigenstates corresponding to eigenenergy $E=0$ are:
      \begin{align}
      \begin{split}
      & |\Psi_{01} \rangle =\frac{1}{\sqrt{|h_{\rm eff}|^2 + \xi_{\q^*}^2}}
                           \begin{bmatrix}
                           h_{\rm eff}   & \xi_{\q^*}  & 0  & 0
                           \end{bmatrix}^T ,\\
      & |\Psi_{02} \rangle =\frac{1}{\sqrt{|h_{\rm eff}|^2 + \xi_{\q^*}^2}}
                             \begin{bmatrix}
                             0 & 0 &h_{\rm eff} & \xi_{\q^*}
                             \end{bmatrix}^T.
      \end{split}
      \end{align}

      The projection operator is then:
      \begin{align}
      \begin{split}
      P &=|\Psi_{01} \rangle \langle \Psi_{01} | + |\Psi_{02} \rangle \langle \Psi_{02} |\\
             &=\frac{1}{\sqrt{|h_{\rm eff}|^2 + \xi_{\q^*}^2}}
                \begin{bmatrix}
                 h_{\rm eff}   & \xi_{\q^*}   & 0         & 0 \\
                 0         & 0            &h_{\rm eff}     & \xi_{\q^*},
                 \end{bmatrix}^T,
      \end{split}
      \end{align}
and the effective Hamiltonian is:
      \begin{eqnarray}\label{eq:effective_hamiltonian}
      H_{2\times 2} &=& P^\dag H_{\bf q}^{\rm MF} P \nonumber\\
      &=&\frac{|h_{\rm eff}|^2}{|h_{\rm eff}|^2 + \xi_{\q^*}^2}
                                   \begin{bmatrix}
                                   \xi_{\q} -\xi_{\q^*}   & i\ {\rm Im} \ S_\bold{q} \\
                                   -i \ {\rm Im} \ S_\bold{q}  &-(\xi_{\q} -\xi_{\q^*})
                                   \end{bmatrix}.\nonumber\\
      \end{eqnarray}
      The basis of the above Hamiltonian $H_{2\times 2} $ are the two degenerate states $|\Psi_{01}\rangle$ and $|\Psi_{02}\rangle$.

      \subsection{1. Mirror symmetric case $J_x=J_y$}
      
      For the case $J_x=J_y=J$, we expand Eq.(\ref{eq:effective_hamiltonian}) near the Weyl points $E=0$ and get:
       \begin{equation}\label{eq:effective}
       H_{2\times 2} \approx \frac{2 J |h_{\rm eff}|^2}{|h_{\rm eff}|^2 + \xi_{\q^*}^2}
                                   \begin{bmatrix}
                                   -\tilde{v}_x \delta q_x  &i \tilde{v}_y \delta q_y \\
                                   -i\tilde{v}_y \delta q_y &\tilde{v}_x \delta q_x
                                   \end{bmatrix},
      \end{equation}
      where $\delta q_x=q_x -q_x^*, \delta q_y=q_y -q_y^*$, and $\tilde{v}_y = \cos q_1^* = \gamma_1$, $\tilde{v}_x= \sin q_1^*=\pm\sqrt{1-\gamma_1^2}$. The $\pm$ sign corresponds to two Weyl points at $E=0$.

To have a nice image of the chirality and winding number of the eigenvectors of the  effective $2\times 2$ Hamiltonian, we rotate the basis  axis for Eq.(\ref{eq:effective}) 
 through the operator $U =e^{-i \frac{\pi}{4} \sigma^y}$.
     The effective Hamiltonian then becomes:
      \begin{equation}\label{eq:effective_symmetric}
       H_{2\times 2} = -  \frac{2 J |h_{\rm eff}|^2}{|h_{\rm eff}|^2 + \xi_{\q^*}^2}
                                   \begin{bmatrix}
                                   0  &\tilde{v}_x \delta q_x -i \tilde{v}_y \delta q_y \\
                                   \tilde{v}_x \delta q_x+i\tilde{v}_y \delta q_y & 0
                                   \end{bmatrix}.
      \end{equation}

      We can do the rescaling $\tilde{q}_x =\tilde{v}_x \delta q_x=\sin q_1^* \delta q_x,\  \tilde{q}_y=\tilde{v}_y \delta q_y=\cos q_1^* \delta q_y $. The effective Hamiltonian then becomes:
     \begin{equation}
      \tilde{H'}_{2\times 2} \propto  - \frac{2 J |h_{\rm eff}|^2}{|h_{\rm eff}|^2 + \xi_{\q^*}^2} \begin{bmatrix}
                                            0 &e^{-i \tilde{\theta}_{\bf q}}\\
                                            e^{i \tilde{\theta}_{\bf q}} &0
                                           \end{bmatrix},
     \end{equation}
     where $\tilde{\theta}_{\bf q}$ is the angle between ${\bf \tilde{q}}=(\tilde{v}_x \delta q_x, \tilde{v}_y \delta q_y)$ and the ${\bf q}_x$ axis. The two eigenvectors of the effective Hamiltonian are two pseudo-spin vectors $|\tilde{\Psi}\rangle_\pm = \frac{1}{\sqrt{2}} ( 1, \ \pm e^{i\tilde{\theta}_{\bf q}})^T$.
     The sign of $\tilde{v_x}$ does not change with the increase of the magnetic field at a given Weyl point, whereas $\gamma_1$ changes sign at $h=h_{c_1}$, corresponding to a sign change of the winding number at the Weyl point. This results in a topological phase transition. 
     
     The same analysis applies to the Weyl points at $E\neq 0$ only with $\gamma_1$ replaced by $\gamma_2=(n_{\chi} + n_{f} -1)J_z/J_x$. Since $\gamma_2 =(n_{\chi} + n_{f} -1)J_z/J_x\leq0$ at $h\leq h_{c_2}$, i.e., it does not change sign at $h<h_{c_2}$, the topological characters of these Weyl points do not change at $h< h_{c_2}$.

   \begin{figure*}[htb]
        \centering
        \includegraphics[width=0.8\textwidth]{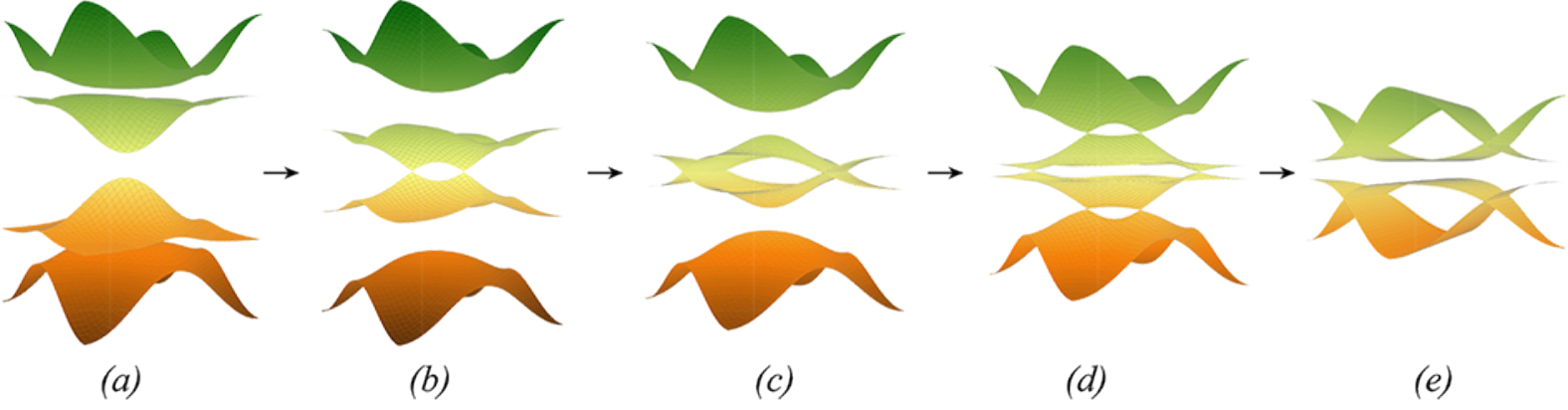}
        \caption{Band structure of  gapped AFM KSL with $ J_x=J_y<0, J_z=2.5J_x$ from the MFT. (a)$h<h_{c_0}$. (b)$h_{c_0}<h<h_{c_1}$. (c)$h=h_{c_1}$. (d)$h_{c_1}<h<h_{c_2}$. (e)$h=h_{c_2}$.}
        \label{fig:gap_afm_band}
     \end{figure*}

 \begin{widetext}   
    \subsection{2. Mirror asymmetric case $J_x \neq J_y$}

     In this case we expand $\xi_\q$ and ${\rm Im} \ S_\bold{q}$ near the Weyl point ${\bf q}^*$ at $E=0$ and get:
     \begin{align}
     \begin{split}
     {\rm Im} \ S_\bold{q} &= J_x \sin q_1 + J_y \sin q_2\\
                           &= J_x \cos q^*_1 \delta q_1+ J_y \cos q^*_2 \delta q_2\\
                           &= (J_x \cos q^*_1 +J_y \cos q^*_2) \delta q_y + (-J_x \cos q^*_1 +J_y \cos q^*_2) \delta q_x\\
                           &= \tilde{\gamma}_1 \delta q_y + \frac{J_y^2 -J_x^2}{\tilde{\gamma}_1} \delta q_x,
     \end{split}
    \end{align}
    
     \begin{align}
     \begin{split}
     {\rm Re} \ S_\bold{q} &=J_z+ J_x \cos q_1 + J_y \cos q_2 \\
                           &= J_z+ J_x \cos q_1^* + J_y \cos q_2^* -J_x \sin q^*_1 \delta q_1-J_y \sin q^*_2 \delta q_2 \\
                           &= {\rm Re} \ S_\bold{q^*} -(J_x \sin q^*_1 +J_y \sin q^*_2) \delta q_y -(-J_x \sin q^*_1 +J_y \sin q^*_2) \delta q_x\\
                           &={\rm Re} \ S_\bold{q^*} + 2 J_x \sin q_1^* \delta q_x.
     \end{split}
    \end{align}
    
The effective $2\times 2$ Hamiltonian becomes:
       \begin{equation}
       H_{2\times 2} \sim \frac{ 2J_x |h_{\rm eff}|^2}{|h_{\rm eff}|^2 + \xi_{\q^*}^2}
                                   \begin{bmatrix}
                                   - \sin q_1^* \delta q_x  &i (\gamma_1 \delta q_y + \frac{J_y^2 -J_x^2}{4\gamma_1J^2_x} \delta q_x)  \\
                                   -i(\gamma_1 \delta q_y + \frac{J_y^2 -J_x^2}{4\gamma_1 J^2_x} \delta q_x) & \sin q_1^* \delta q_x
                                   \end{bmatrix}.
      \end{equation}

     After the same rotation of the basis axis as for the mirror symmetric case in the last subsection, the effective Hamiltonian becomes:
      \begin{equation}\label{eq:effective_asymmetric}
       H'_{2\times 2} =-\frac{ 2J_x |h_{\rm eff}|^2}{|h_{\rm eff}|^2 + \xi_{\q^*}^2}
                                   \begin{bmatrix}
                                   0  & \sin q_1^* \delta q_x -i (\gamma_1 \delta q_y + \frac{J_y^2 -J_x^2}{4\gamma_1 J^2_x} \delta q_x)  \\
                                    \sin q_1^* \delta q_x +i (\gamma_1 \delta q_y + \frac{J_y^2 -J_x^2}{4\gamma_1 J^2_x} \delta q_x) & 0
                                   \end{bmatrix}.
      \end{equation}
      Compared to the effective Hamiltonian Eq. (\ref{eq:effective_symmetric}) for the mirror symmetric case, Eq.(\ref{eq:effective_asymmetric}) is equivalent to Eq. (\ref{eq:effective_symmetric}) by an rotation of axis $ {\bf q}_y$ while keeping the ${\bf q}_x$ axis still.

  \end{widetext}        
      
       The two eigenvectors of the effective Hamiltonian still has the form $|\tilde{\Psi}\rangle_\pm = \frac{1}{\sqrt{2}} ( 1, \ \pm e^{i\tilde{\theta}_{\bf q}})^T$, where $\tilde{\theta}_{\bf q}$ now is the angle between ${\bf \tilde{q}}=(\sin q_1^*\delta q_x, \gamma_1 \delta q_y + \frac{J_y^2 -J_x^2}{4\gamma_1 J^2_x} \delta q_x )$ and the ${\bf q}_x$ axis.  The winding number of these eigenvectors  is still determined by the sign of $\sin q_1^*$ and $\gamma_1$.

     For the asymmetric AFM gapless case $J_y=0.95J_x, J_x=J_z<0$ in section III.2 in the main text, the gapless regime at $h<h'_{c_1}$ has $\gamma_1<0$ and the the gapless regime at $h''_{c_1}<h<h_{c_2}$ has $\gamma_1>0$. For the reason, for a given Weyl point, the winding number  at $h<h'_{c_1}$ and $h''_{c_1}<h<h_{c_2}$ has opposite sign. The winding number at two different Weyl points also has opposite sign at given $h$ since $ \sin q_1^*$ has opposite sign at the two Weyl points.

  \section{II. The mean field results for the gapped AFM case with $J_x=J_y$}
         \begin{figure}[!htb]
        \centering
        \includegraphics[width=8cm]{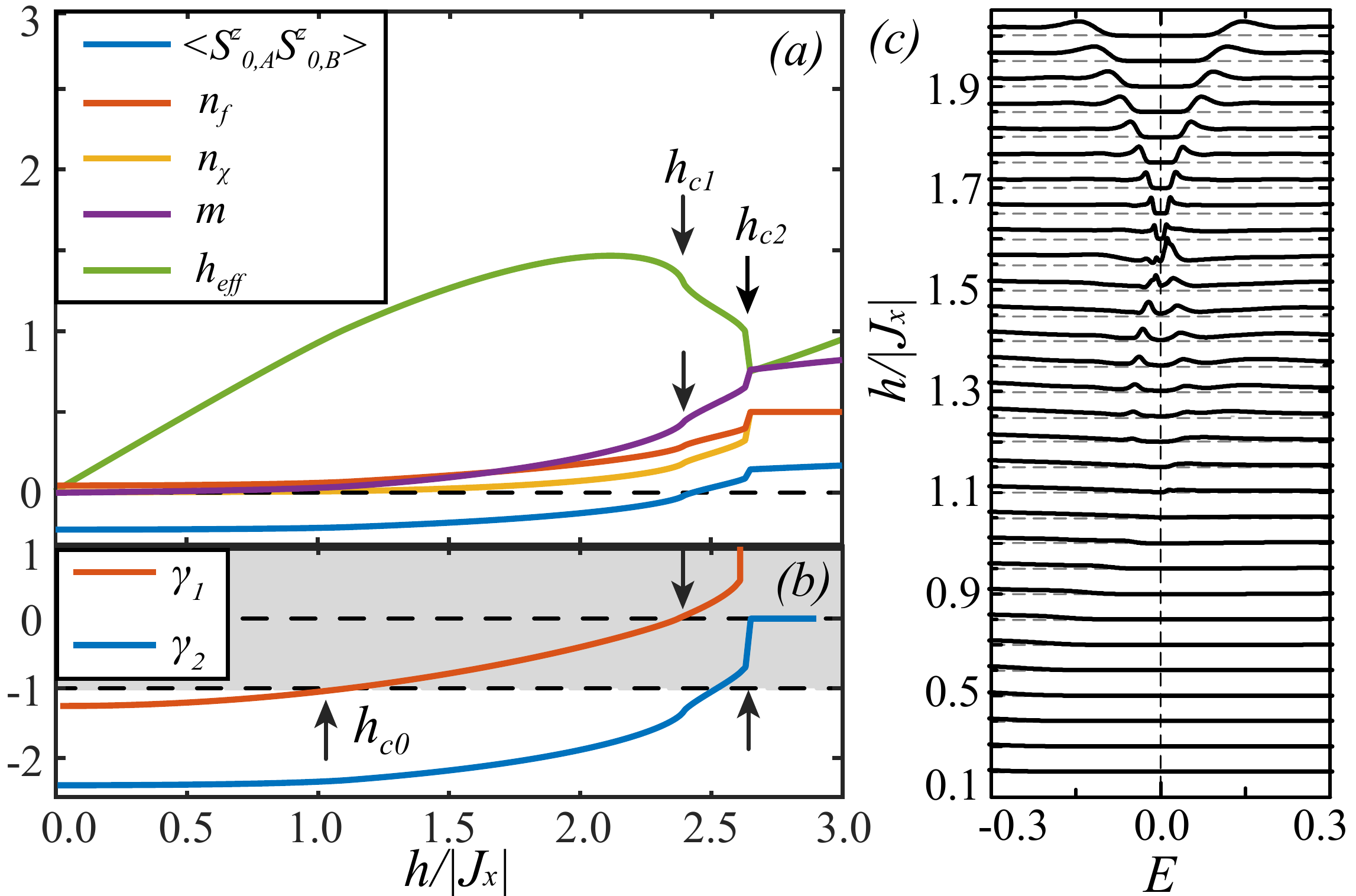}
         \caption{(a)Mean field parameters for gapped AFM KSL with $ J_x=J_y<0, J_z=2.5J_x$ as a function of magnetic field $h$. (b)$\gamma_1$ and $\gamma_2$ as a function of $h$. (c)The low energy density of states of matter fermion quasiparticles obtained from DMFT. The labels of the curves are the same as Fig.\ref{fig:AFM_results}(c).}
        \label{fig:gap_afm_para}
   \end{figure}

      In this section, we  show the mean field results for a typical  gapped AFM case with $ J_x=J_y<0, J_z=2.5J_x$.

     Fig.\ref{fig:gap_afm_band} shows the evolution of the band structures of the case with the increase of magnetic field $h$. At $h=0$, the bands are gapped at $E=0$. With the increase of $h$, at some critical value $h_{c_0}$, the bands become gapless at $E=0$ with two nodal points as shown in Fig.\ref{fig:gap_afm_band}(b). As $h$ further increases and reaches a critical field $h_{c_1}$, two nodal lines appear at $E=0$ as shown in Fig.\ref{fig:gap_afm_band}(c), corresponding to a topological phase transition. At $h>h_{c_1}$, the nodal lines turn to nodal points again as shown in Fig.\ref{fig:gap_afm_band}(d) until at $h=h_{c_2}$, a gap opens at $E=0$ and the system turns to the trivial polarized state shown in Fig.\ref{fig:gap_afm_band}(e). This is consistent with the analysis for the gapped AFM case with mirror symmetry in the main text.

     Fig.\ref{fig:gap_afm_para}(a) shows the mean field parameters as a function of $h$ for the same case. It demonstrates the phase transitions at $h_{c_1}$ and $h_{c_2}$ clearly whereas  the transition from the gapped phase to gapless phase at $h_{c_0}$ is continuous.   The evolution of $\gamma_1$ and $\gamma_2$ with the magnetic field shown in Fig.\ref{fig:gap_afm_para}(b) is consistent with the band structures in Fig.\ref{fig:gap_afm_band}.
     
     The DMFT results of the low energy density of states of the matter fermion quasiparticles for this case are shown in Fig. \ref{fig:gap_afm_para}(d). We can see that the evolution of the low energy density of states with the magnetic field from the DMFT agrees qualitatively with the band structure shown in Fig.\ref{fig:gap_afm_band} from the MFT.


\begin{thebibliography}{99}
\bibitem{Footnote1}To eliminate the $i$ factor explicitly in the Hamiltonian, we choose the gauge for $\chi_{\bf r}$ in its definition such that it transform as $\chi_{\bf r} \to -\chi_{\bf r}$ under TRS whereas $f_{\bf r}$ transform as $f_{\bf r}\to f_{\bf r}$.
\bibitem{Footnote0}This is natural if the system remains a spin liquid.
\end{thebibliography}
\end{document}